\documentclass[11pt]{article}          %% 
\usepackage{axodraw}

\def\vec#1{\mbox{\boldmath $#1$}}

\title{\centerline{\normalsize SINP/TNP/02-25 \hfill hep-th/0209012}
\bigskip 
\bf A non-singular potential for the Dirac monopole} 

\author{
{\bf Ranjan Kumar Ghosh}\thanks{rkg\_1978@yahoo.com}\\
\normalsize Bidhannagar College, EB-2 Salt Lake City, 
Calcutta 700064, India
\and 
{\bf Palash B. Pal\thanks{pbpal@theory.saha.ernet.in}}\\ 
\normalsize Saha Institute of Nuclear Physics, 1/AF Bidhan-Nagar, 
Calcutta 700064, India}

\date{August 2002}

\begin{document}

\maketitle 

%%%%%%%%%%%%%%%%%%%%%%%%% 
\begin{abstract} \noindent\small
We propose a new vector potential for the Abelian magnetic monopole.
The potential is non-singular in the entire region around the
monopole.  We argue how the Dirac quantization condition can be
derived for any choice of potential.

\end{abstract} 
%%%%%%%%%%%%%%%%%%%%%%%%% 

Magnetic monopoles have not been found experimentally, but there
are strong theoretical reasons to believe that they exist.  Dirac
\cite{dirac} showed that if a magnetic monopole exists, its magnetic
charge $g$ obeys the relation
\begin{eqnarray}
qg = n/2 
\label{qcond}
\end{eqnarray}
in natural units, where $q$ is the charge of any particle and $n$ is
an integer.  This implies quantization of electric charges.

The magnetic field of a monopole is given by
\begin{eqnarray}
\vec B = {g \over r^2} \; \hat r \,,
\label{B}
\end{eqnarray}
where $r$ is the radial distance from the monopole and $\hat r$ is the
unit vector in the radial direction.  Thus, the integral of the
magnetic flux on a closed surface containing the monopole does not
vanish: 
\begin{eqnarray}
\oint d\vec S \cdot \vec B = 4\pi g \,.
\label{totflux}
\end{eqnarray}
Accordingly, one cannot define a vector potential globally by the
relation 
\begin{eqnarray}
\vec B = \vec\nabla \times \vec A \,,
%\label{}
\end{eqnarray}
since that would imply vanishing of the divergence of $\vec B$, and
consequently of the surface integral on the left side of Eq.\
(\ref{totflux}).  It may not be a problem classically, where one could
work entirely in terms of the electric and magnetic fields.  But in
quantum theory, $\vec A$ plays a fundamental role, and so it must be
defined.

Dirac \cite{dirac} circumvented this problem by hypothesizing a
thin string which carries all the flux and definining the
divergence-free $\vec A$ everywhere else.  This Dirac string must be
unphysical and therefore undetectable.  This leads to the Dirac
quantization formula \cite{dirac,saha,wilson,Coleman:1982cx} of Eq.\
(\ref{qcond}).  Much later, Wu and Yang \cite{Wu:es,Wu:ge} showed that
the Dirac string can be totally avoided in a formulation where one
uses two different patches for the vector potential for the space
around a monopole:
\begin{eqnarray}
A_r = A_\theta = 0 \,, \qquad 
A_\phi = \cases{
{\displaystyle g\over \displaystyle r\sin\theta} \,  (1 -\cos\theta) 
& for $0\leq \theta\leq \frac12\pi+\epsilon$, \cr \cr
{\displaystyle g\over \displaystyle r\sin\theta} \,  (-1 -\cos\theta) 
& for $\frac12\pi-\epsilon \leq \theta\leq\pi$,}
\label{Wu-Yang}
\end{eqnarray}
for any arbitrary $\epsilon$ in the range $0<\epsilon<\frac12\pi$.
Notice each patch has a singularity if we try to extend them over the
entire region around the monopole as Dirac did, but is regular in its
restricted domain of definition shown above.  In the overlap region
$\frac12\pi-\epsilon \leq \theta\leq \frac12\pi+\epsilon$, the two
patches are related by a gauge transformation:
\begin{eqnarray}
\vec A^{(+)} = \vec A^{(-)} + \vec\nabla (2g\phi) \,,
%\label{}
\end{eqnarray}
where the superscripted plus and minus sign denotes the patch which
covers the upper and lower hemisphere respectively.  On wave functions
(for quantum mechanics) or on other fields (for quantum field theory)
in the theory, there will be a corresponding transformation of the
form $\exp(2i\phi qg)$.  Demanding that this is single valued
everywhere, Wu and Yang \cite{Wu:es} obtained the Dirac quantization
condition, Eq.\ (\ref{qcond}).

In the same paper \cite{Wu:es}, Wu and Yang argued that it is
impossible to define a singularity-free potential for the space around
the monopole.  Their argument starts with the assumption that such a
potential exists.  The loop integral $\oint d\vec x\cdot \vec A$
around any closed curve must then give the magnetic flux through the
surface enclosed.  If we keep $r$ and $\theta$ constant and traverse a
loop in the $\phi$-direction, the magnetic flux through such a polar
cap would be given by
\begin{eqnarray}
\Phi(r,\theta)=2\pi g(1-\cos\theta) \,.
\label{fluxtheta}
\end{eqnarray}
This gives $\Phi(r,\pi)=4\pi g$.  Wu and Yang comment that this is a
contradiction, since at $\theta=\pi$ the curve has shrunk to a point
and therefore the flux through it must be zero.  This proves, reductio
ad absurdum, that a non-singular potential cannot exist, according to
Wu and Yang~\cite{Wu:es}.

In fact, there is no contradiction here.  Any closed curve on the
surface of a sphere is the boundary of two regions.  For example, the
equator can be thought of as the boundary of the northern hemisphere
or of the southern hemisphere.  Likewise, the closed ``curve'' at
$\theta=\pi$ is the boundary of either a region of zero area, or of
the entire remaining area.  Eq.\ (\ref{fluxtheta}) gives the flux
through the region bounded by the curve which contains the point
$\theta=0$.  This is the surface of the entire sphere, so it is no
wonder that the flux through that would come out to be $4\pi g$.

There is therefore no argument to show that one cannot define a
non-singular potential for a monopole.  And in fact, non-singular
potentials exist.  Here we propose such a potential:
\begin{eqnarray}
A_r =0 \,, \qquad A_\theta = -\, {g\over r} \, \phi \sin\theta  ,
\qquad A_\phi = 0 \,.
\label{ourA}
\end{eqnarray}
It is trivial to check that the curl of this potential gives the
magnetic field of Eq.\ (\ref{B}).  It is also quite obvious that,
apart from the essential singularity at $r=0$, there is no other
singularity of this potential.  The Wu-Yang patches, on the other
hand, have singularities outside their domain of definition, which is
why no patch can be extended to the entire space.  For example, $\vec
A^{(+)}$ is singular on the entire half-line $\theta=\pi$, whereas
$\vec A^{(-)}$ is singular on the half-line $\theta=0$.

Although our potential is multi-valued because the azimuthal angle
$\phi$ is defined only modulo $2\pi$, this does not cause any problem.
The reason is that, the value of the potential for two values of
$\phi$ separated by $2\pi$ are related by a gauge transformation:
\begin{eqnarray}
\vec A(r,\theta,\phi+2\pi) = \vec A(r,\theta,\phi) + \vec\nabla (2\pi
g\cos\theta) \,.
%\label{}
\end{eqnarray}
Further, our potential is very simply related to the Wu-Yang patches
through gauge transformation:
\begin{eqnarray}
\vec A^{(\pm)} = \vec A + \vec\nabla \Big( (\pm1 - \cos\theta) \phi
\Big) \,.
%\label{}
\end{eqnarray}
Thus, the two potentials should have identical physical implications,
including the quantization condition.  However, we should be able to
derive the quantization condition without making any reference to the
Wu-Yang potential, which is what we argue below.

To begin with, we make an important point about the Wu-Yang derivation
of the quantization condition, which we have outlined above.  The
proof relies on the gauge transformation connecting the different
patches.  However, it should be realized that such ``sewing conditions
for patches'' cannot guarantee the correct result.  To illustrate the
point, let us suppose that, instead of the two patches of Eq.\
(\ref{Wu-Yang}) used by Wu and Yang, we take three different patches
defined by
\begin{eqnarray}
A_r = A_\theta = 0 \,, \qquad 
A_\phi = \cases{
{\displaystyle g\over \displaystyle r\sin\theta} \,  (1 -\cos\theta) 
& for $0\leq \theta < \frac13\pi$, \cr \cr
{\displaystyle g\over \displaystyle r\sin\theta} \,  (a-\cos\theta) 
& for $\frac14\pi < \theta < \frac34\pi$, \cr \cr
{\displaystyle g\over \displaystyle r\sin\theta} \,  (-1 -\cos\theta) 
& for $\frac23\pi < \theta\leq\pi$,}
\label{3patches}
\end{eqnarray}
where $a\neq\pm1$.  In this case, the sewing condition between the two
patches in the region $\frac14\pi < \theta < \frac13\pi$ will give
$qg=n/(1-a)$, which is not the Dirac quantization condition.  One
might then make the further stipulation that one must take only the
minimum number of patches necessary to cover the space around the
monopole.  But this only shows that a proper derivation of the
quantization condition must somehow take into account the global
properties of the space around the monopole, not any local condition.

This can be done, and the quantization condition derived, without
making any reference to any specific form of the potential.  All one
needs is the property, emphasized by Wu and Yang \cite{Wu:es}, that a
complete and unsuperfluous description of electromagnetism is provided
by the phase factors
\begin{eqnarray}
\exp\Big(iq\oint dx^\mu A_\mu\Big)
%\label{}
\end{eqnarray}
around all possible closed loops.  For a purely magnetic field,
$A_0=0$, so the integral in the exponent becomes the line integral of
the vector potential around a closed loop.

For the space around the monopole, consider such a closed loop.  To be
specific, one can consider a loop on a sphere of radius $r$, although
this restriction is not essential.  No matter which vector potential
we use, the line integral around a loop $C$ will equal the flux of
magnetic field through a surface $S$ whose boundary is $C$.

It is not clear though what we mean by the surface $S$.  As pointed
out earlier, any loop on a sphere is the boundary of two complementary
surfaces.  If the outward flux through one of these surfaces is
$\Phi$, the outward flux through the other must be $4\pi g-\Phi$.  The
direction in which the line integral on $C$ is taken determines the
direction in which we have to consider the normal to the surface $S$.
If the direction of the line integral is such that on the first
surface the normals are outward, they will be inward for the second
surface.  To apply Stokes' theorem, we must then use the inward flux
for the second surface, which is $\Phi-4\pi g$.  Thus, Wu and Yang's
phase factor is given by two expressions, which must give the same
result:
\begin{eqnarray}
\exp \Big( iq \Phi \Big) = \exp \Big( iq (\Phi - 4\pi g) \Big) \,.
%\label{}
\end{eqnarray}
This implies
\begin{eqnarray}
e^{4\pi iqg} = 1 \,,
%\label{}
\end{eqnarray}
which gives the Dirac quantization condition.  As indicated earlier,
this makes no reference to the specific form of the vector potential,
and in particular applies to our potential as well.

It might naively seem that this argument faces a problem if, for our
potential of Eq.\ (\ref{ourA}), we consider a loop with constant
$\theta$.  Since our potential does not have any component in the
$\phi$-direction, it might seem that the line integral on this loop
would vanish.  The same problem would occur for the Wu-Yang potential
if we consider a loop consisting of two semi-circles of constant
$\phi$, as shown in Fig.~\ref{f:WYloop}a. On the other hand, the
magnetic field lines are radial and isotropic.  So, through any loop
subtending a solid angle $\Omega$ at the center, the magnetic flux
should be $g\Omega$.  This poses an apparent paradox.  Let us first
discuss how this paradox is resolved in the Wu-Yang case, before
commenting on our potential.

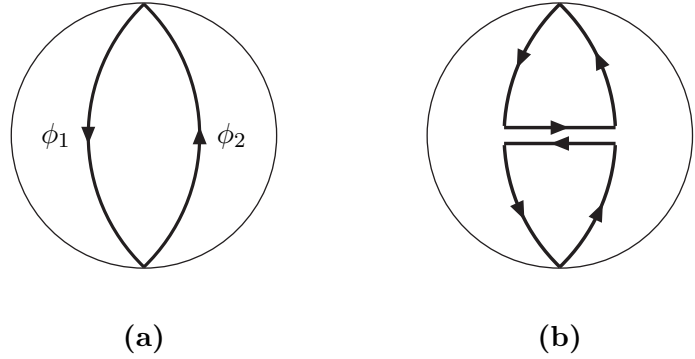
\begin{figure}
\begin{center}
\begin{picture}(60,60)(-30,-30)
\CArc(0,0)(50,0,360)
\Text(-13,0)[c]{$\phi_1$}
\Text(13,0)[c]{$\phi_2$}
\SetWidth{1.2}
\ArrowArc(48,0)(69,134,226)
\ArrowArc(-48,0)(69,314,46)
\Text(0,-30)[c]{\bf (a)}
\end{picture}
\begin{picture}(60,60)(-30,-30)
\CArc(0,0)(50,0,360)
\SetWidth{1.3}
\ArrowArc(48,0)(69,134,177)
\ArrowArc(48,0)(69,183,226)
\ArrowLine(-21,3)(21,3)
\ArrowArc(-48,0)(69,314,357)
\ArrowArc(-48,0)(69,3,46)
\ArrowLine(21,-3)(-21,-3)
\Text(0,-30)[c]{\bf (b)}
\end{picture}
\end{center}
\caption{Line integral of the Wu-Yang potential is naively zero around
the loop on the sphere shown in part (a) of the figure.  However, as
argued in the text, the loop should be decomposed into two
parts as shown in part (b) in order to calculate the line integral
properly.}\label{f:WYloop}
\end{figure}
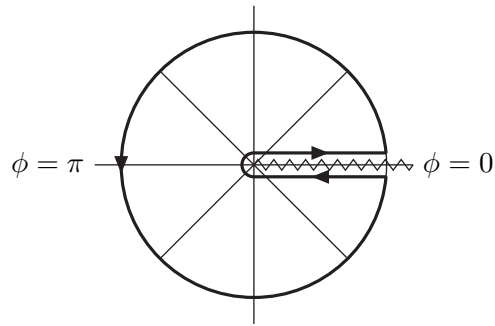
\begin{figure}
\begin{center}
\begin{picture}(100,60)(-50,-30)
\CArc(0,0)(50,0,360)
\Line(0,0)(0,60)
\ZigZag(0,0)(60,0)2{10}
\Line(0,0)(0,-60)
\Line(0,0)(60,0)
\Line(0,0)(-60,0)
\Line(0,0)(35,35)
\Line(0,0)(35,-35)
\Line(0,0)(-35,35)
\Line(0,0)(-35,-35)
\Text(25,0)[l]{$\phi=0$}
\Text(-25,0)[r]{$\phi=\pi$}
\SetWidth{1.2}
\ArrowArc(0,0)(50,5,355)
\CArc(0,0)(4.5,90,270)
\ArrowLine(0,4.5)(50,4.5)
\ArrowLine(50,-4.5)(0,-4.5)
\end{picture}
\end{center}
\caption{The figure shows a sphere in the polar projection. The center
represents the north pole. The radial lines are lines of constant
$\phi$, whereas the circular line is a line of constant
$\theta$. If one wants to perform the line integral of our vector
potential over a loop of constant $\theta$, the loop will have to be
deformed as shown in the figure.}\label{f:ourloop}
\end{figure}
The crucial thing to realize is that the loop in Fig.~\ref{f:WYloop}a
crosses discontinuities of the Wu-Yang potential.  So, one should be
careful while taking the line integral.  The correct way of doing this
is to add two equal and opposite lines in the middle, shown as
separate lines for the sake of clarity in Fig.~\ref{f:WYloop}b.  These
lines break the loop into two parts.  Each part of the loop entirely
lies in a region in which there is a single continuous potential.  For
each part, the line integral can be calculated in a straight forward
way.  Then one should add the contributions to obtain the line
integral over the entire loop.  If the two semicircles of the loop are
at $\phi=\phi_1$ and $\phi=\phi_2$, the line integral on the upper
part of the loop is given by
\begin{eqnarray}
\Phi_{(+)} = \int_{\phi_1}^{\phi_2} d\phi \; r\sin\theta
A^{(+)}_\phi \,,
%\label{}
\end{eqnarray}
whereas on the lower part it is given by
\begin{eqnarray}
\Phi_{(-)} = \int_{\phi_2}^{\phi_1} d\phi \; r\sin\theta
A^{(-)}_\phi \,.
%\label{}
\end{eqnarray}
Using the form of the potential from Eq.\ (\ref{Wu-Yang}), we then
obtain the total flux through the closed loop to be
\begin{eqnarray}
\Phi = 2g (\phi_2-\phi_1) \,.
%\label{}
\end{eqnarray}
This is the correct value from the isotropy argument. When
$\phi_2-\phi_1=2\pi$, it gives the total flux to be $4\pi g$.

We now consider a similar situation with our potential.  The
contentious loops in this case are lines of constant $\theta$.  Since
$A_\phi$ vanishes for our potential, it would naively seem that the
line integral along such a loop vanishes as well.  The important thing
is to recall that our potential of Eq.\ (\ref{ourA}) is multi-valued.
One can circumvent this problem by making the substitution
\begin{eqnarray}
\phi \to \phi \mathop{\rm mod} 2\pi 
\label{mod}
\end{eqnarray}
in Eq.\ (\ref{ourA}), which does not affect the curl of the potential.
However, the potential is now discontinuous at $\phi=0$, and the
contentious loops certainly cross this discontinuity.  As in the
Wu-Yang case, we now have to avoid the discontinuity by going around
the north pole, as shown in Fig.~\ref{f:ourloop}.  The entire
contribution to the line integral now comes from the added pair of
lines, and is given by
\begin{eqnarray}
\int_0^\theta d\theta \; r \Big( A_\theta(0+) - A_\theta(0-) \Big) \,, 
%\label{}
\end{eqnarray}
where $A_\theta(0+)$, for example, denotes the value of $A_\theta$ for
a slightly positive value of $\phi$.  Using Eq.\ (\ref{ourA}), we now
obtain Eq.\ (\ref{fluxtheta}), which is the correct flux through such
a north polar cap.  If instead we avoid the disconinuity by going
around the south pole, we would obtain the flux through the
complementary surface.

We have thus provided a vector potential for the magnetic monopole
which is non-singular everywhere in the space around the monopole.  It
is admittedly multi-valued.  It can be made single-valued by
restricting the azimuthal angle as in Eq.\ (\ref{mod}). It then
becomes discontinuous, but the whole space can nevertheless be covered
with a single patch.  Because of this feature, we hope our potential
will be easier to work with in practical situations.

\paragraph*{Note added~: } After submitting this paper to the hep-th
archive, we came to know through Douglas Singleton that Eq.\
(\ref{ourA}) appears in Ref.~\cite{arfken}.  The related physics
issues have no mention there.

%%%%%%%%%%%%%%%%%%%%


\begin{thebibliography}{[99]}
%
\bibitem{dirac} P.~A.~M. Dirac, Proc. Roy. Soc. London A133, 60
(1931).

\bibitem{saha} M.~N. Saha, Ind. J. Phys. 10, 141 (1936);
Phys. Rev. 75, 1968 (1949).

\bibitem{wilson} H.~A. Wilson, Phys. Rev. 75, 309 (1949).

%\cite{Coleman:1982cx}
\bibitem{Coleman:1982cx}
S.~R.~Coleman,
%``The Magnetic Monopole Fifty Years Later,''
in ``Les Houches 1981, Proceedings, Gauge Theories In High
    Energy Physics, Part 1'', pp 461-552.

%\cite{Wu:es}
\bibitem{Wu:es}
T.~T.~Wu and C.~N.~Yang,
%``Concept Of Nonintegrable Phase Factors And Global Formulation Of Gauge  Fields,''
Phys.\ Rev.\ D {\bf 12}, 3845 (1975).
%%CITATION = PHRVA,D12,3845;%%


%\cite{Wu:ge}
\bibitem{Wu:ge}
T.~T.~Wu and C.~N.~Yang,
%``Dirac Monopole Without Strings: Monopole Harmonics,''
Nucl.\ Phys.\ B {\bf 107}, 365 (1976).
%%CITATION = NUPHA,B107,365;%%

\bibitem{arfken} 
G.~B. Arfken, H-J. Weber: ``Mathematical Methods for
Physicists'' (Harcourt/Academic Press), 5th Edition, page 130.


\end{thebibliography}
\end{document}